\documentclass[aps,twocolumn,prl,amsmath,amssymb,floatfix,superscriptaddress]{revtex4-1}
\usepackage[english]{babel}
\usepackage{latexsym,amsmath,amssymb}
\usepackage{graphicx}
\usepackage{esint}
\usepackage{braket}
\usepackage{multirow}
\usepackage{dcolumn}
\usepackage{color}
\usepackage{physics}
\usepackage{tikz}
\usepackage{subcaption}
\usepackage{siunitx}
\usepackage{bm}
\usepackage[mathlines]{lineno}
\usepackage{hyperref}
\usepackage{epsfig}
\usepackage{epstopdf}
\usepackage{lipsum}
\begin{document}
\title{Precision measurements with cold atoms and trapped ions}

\author{Qiuxin Zhang}
\affiliation{Department of Physics, Renmin University of China, Beijing 100872, China}
\author{Yirong Wang}
\affiliation{Department of Physics, Renmin University of China, Beijing 100872, China}
\author{Chenhao Zhu}
\affiliation{Department of Physics, Renmin University of China, Beijing 100872, China}
\author{Yuxin Wang}
\affiliation{Department of Physics, Renmin University of China, Beijing 100872, China}

\author{Xiang Zhang}
\affiliation{Department of Physics, Renmin University of China, Beijing 100872, China}
\author{Kuiyi Gao}
\thanks{kuiyi@physik.uni-bonn.de}
\affiliation{Department of Physics, Renmin University of China, Beijing 100872, China}

\author{Wei Zhang}
\thanks{wzhangl@ruc.edu.cn}
\affiliation{Department of Physics, Renmin University of China, Beijing 100872, China}
\affiliation{Beijing Key Laboratory of Opto-electronic Functional Materials and Micro-nano Devices, Renmin University of China, Beijing 100872,China}
	
\begin{abstract}
	Recent progresses on quantum control of cold atoms and trapped ions in both the scientific and technological aspects greatly advance the applications in precision measurement. Thanks to the exceptional controllability and versatility of these massive quantum systems, unprecedented sensitivity has been achieved in clocks, magnetometers and interferometers based on cold atoms and ions. Besides, these systems also feature many characteristics that can be employed to facilitate the applications in different scenarios. In this review, we briefly introduce the principles of optical clocks,  cold atom magnetometers and atom interferometers used for precision measurement of time, magnetic field, and inertial forces. The main content is then devoted to summarize some recent experimental and theoretical progresses in these three applications, with special attention being paid to the new designs and possibilities towards better performance. The purpose of this review is by no means to give a complete overview of all important works in this fast developing field, but to draw a rough sketch about the frontiers and show the fascinating future lying ahead. 
\end{abstract} 
	
	

\maketitle
	
	\section{Introduction}
	\label{sec:intro}
	
	Physics, as the fundamental discipline of natural sciences, is a reasonable enterprise that can provide us with knowledge of the physical world. It is based on valid experimental evidence, rational discussion, and mathematical derivation. Among these methodologies, experiment plays a vital role both to discover new phenomenon that is in need of explanation, and to exhibit scientific facts that can test various theories. Although the textbooks at different levels are often aiming at the explanation of successful theories, along the history of physics it is truly the experiments that mark the breakthrough. In many cases, a quantitative measurement with unprecedented precision is the most, and maybe the only, legitimate route to open a new era of our understanding of the world.
	
	One remarkable example of the power of measurement is the discovery of Kepler's law of planetary motion. If Tycho has not been able to locate objects in the sky within an uncertainty of only a few minutes, improved by nearly an order of magnitude from previous data, Kepler would not find any discrepancy of an orbit composed of circles or ovoids. Finally, Kepler concluded that planets, including the Earth, move on ellipses, with the Sun at a focus, but not the center. In fact, this solution is the only choice that can reconcile Tycho's precise data. Kepler's finding not only provides a solid evidence in support of Copernican theory over Ptolemy, but also shines new light to our understanding of the nature of planetary motion. As an ellipse cannot be obtained by any combination of rotary motion, the long-lasting picture raised by ancient Greeks of planets residing on rotating spheres must be abandoned. It is then natural to accept that the planet orbits are empty, and the planets are moving under the constraint of a force that can exert on them by the Sun remotely. 
	
	The success of theory of relativity is another great triumph of measurement. By inventing one of the most accurate equipments in human history, Michelson and Morley ruled out the possibility of Ether, which, if exists at all, should cause a displacement one order of magnitude bigger than the equipment accuracy. Astronomers consistently recorded the position of the Mercury over years and found an unreasonable discrepancy of 43 seconds per century in the orbit precession. Eddington performed an extraordinary measurement during a solar eclipse, and obtained results in quantitative consistent with Einstein's new theory. And most recently in 2016, the LIGO project received the first direct signal of gravitational waves from two merging blackholes billions of light years away, witnessing the prediction made by Einstein more than a hundred years ago. 
	
	Like the two aforementioned examples, the frontiers of physics at this moment are also waiting for the ground-breaking results from measurement. The detection of electronic dipole moment, high energy cosmic particles, gravitational waves, difference between gravitational mass and inertial mass, and possible candidates of dark matter particles, are all bold explorations which can widely open our eyes and lead us to regimes where no one has gone before. Besides, measurement with higher accuracy also has versatile applications in state-of-the-art technology, including navigation of unpersonalized objects, monitoring the ultrafast dynamics of a chemical reaction, and probing tiny structures inside living cells, to name a few.  
	
	On one hand, the fundamental principles of quantum mechanics are mostly about measurement. It is literally impossible to understand the true nature of measurement without the knowledge about operators, eigenstates, and uncertainty principle. In another point of view, measurement is a process of information transformation, and information itself is physical and has to obey quantum mechanics. On the other hand, the advances in quantum technology provide us a versatile toolbox of manipulating microscopic objects, which can be implemented to extract tiny disturbance in a short time interval at a localized region. Thus, an exciting direction about measurement emerged and got boosted in the past few decades, known as quantum sensing or quantum metrology.
	
	There are many different ways to build a classical device to finish a measurement of certain physical quantities. Imagine one uses a caliper for precision measurement of the sizes of different objects. It is desirable to make a caliper from a material with extremely low coefficient of expansion, as a very stable reference of length. On the contrary, one would choose a material to build an expansion thermometer with extremely high coefficient of expansion, so that a high sensitivity of temperature measurement is possible. In some differential measurement for even higher accuracy, measuring the same quantities by two devices simultaneously and comparing their results is a great solution. In the field of quantum sensing and quantum metrology, we also take the same basic and useful approaches. 
	
	In this short review, we briefly introduce some principles, recent progresses, perspectives and challenges in the field of precision measurement using cold atoms or trapped ions. The overall plan of the paper is as follows. In Sec.~\ref{sec:time}~~ we present some basic principles and recent advances about time measurement. The topic of Sec.~\ref{sec:Bfield}~~ is about measurement of magnetic field with cold atom magnetometers. In Sec.~\ref{sec:force}, we discuss the advances and applications in inertial forces measurement with cold atoms. The reader will notice that most of the developments discussed in this review occurred after 2010. Our intention in reviewing these recent efforts is to provide an idea of the current directions and frontiers related to precision measurement with cold atoms and trapped ions. This review is prepared for advanced graduate students, post-docs, and colleagues working in the field of precision measurement. A general audience whose expertise is not in this field may refer to some tutorial reviews or books~~\cite{nawrocki-book, siegner-book, degen-review}. The topics and the papers that are included in this review are organized in such a way to give a reasonably fluent presentation. We did not try to assign credits or priorities in the order of content. If any of the readers feel that their contributions were not properly acknowledged or reviewed, we would ask them to attribute our errors and omissions to our own stupidity, ignorance, laziness and haste rather than any malicious behavior.
	
	\section{Time measurement}
	\label{sec:time}
	
	In the revised definition of International System of Units (SI), announced by the 26th Conférence Générale des Poids et Mesures (CGPM) on May 20, 2019, all fundamental SI units except Mole are directly or indirectly linked to Second. This is because the measurement accuracy of time is the highest among the seven SI units, which ensures the long-term stability and global versatility of the SI system to meet the consistent need of precision measurement and in academic researches and technical applications.
	
	Over the years, applications in multidisciplinary fields require more precise time measurement and synchronization, including navigation systems~\cite{dow2009international}, telecommunications, very long baseline interferometry (VLBI) telescope~\cite{normile2011first} and fundamental physics~\cite{rosenband2008frequency}. The key of a good clock is an absolute reference frequency which can be obtained in our daily life by, e.g., the swing of a pendulum or the voltage-driven oscillations of a quartz crystal. However, pendulum and quartz crystal are usually susceptible to environmental disturbances, such as temperature fluctuations. Nowadays, the best clocks are those that choose certain transitions of atoms as the reference frequency~\cite{ludlow2015optical}. The current definition of ``second'' is based on a microwave clock with a unperturbed ground state hyperfine transition frequency of \SI{9.19}{\giga\hertz} in $^{133}$Cs atoms, which is just a more strict definition in 2018 than in 1967 without improvement in frequency accuracy.
	
	However, the state-of-the-art Cs atom clock~\cite{parker2009long} is approaching its practical limitations. For more accurate timing, the search for clocks based on optical transition has become a hot research area in the past decades. Studies in this direction is boosted by the advent of two key technologies. Firstly, with the invention of femtosecond-laser optical frequency combs, the measurement of the microwave band extends to visible light frequency~\cite{reichert1999measuring, diddams1999broadband}. Secondly, significant progresses have been made in the fields of atomic and ion manipulation and precise optical frequency control, leading to a higher frequency accuracy~\cite{wieman1999atom}. Now, the frequency accuracy of optical clocks have reached the range of 10$^{-18}$, which is three orders more accurate than a microwave clock. Thus, in 2016, Consultative Committee for Time and Frequency gave a roadmap for modifying the definition of Second based on optical transitions around 2025.
	
	\subsection{Principles}
	
\begin{figure*}[t]
\centering
\includegraphics[scale=0.45]{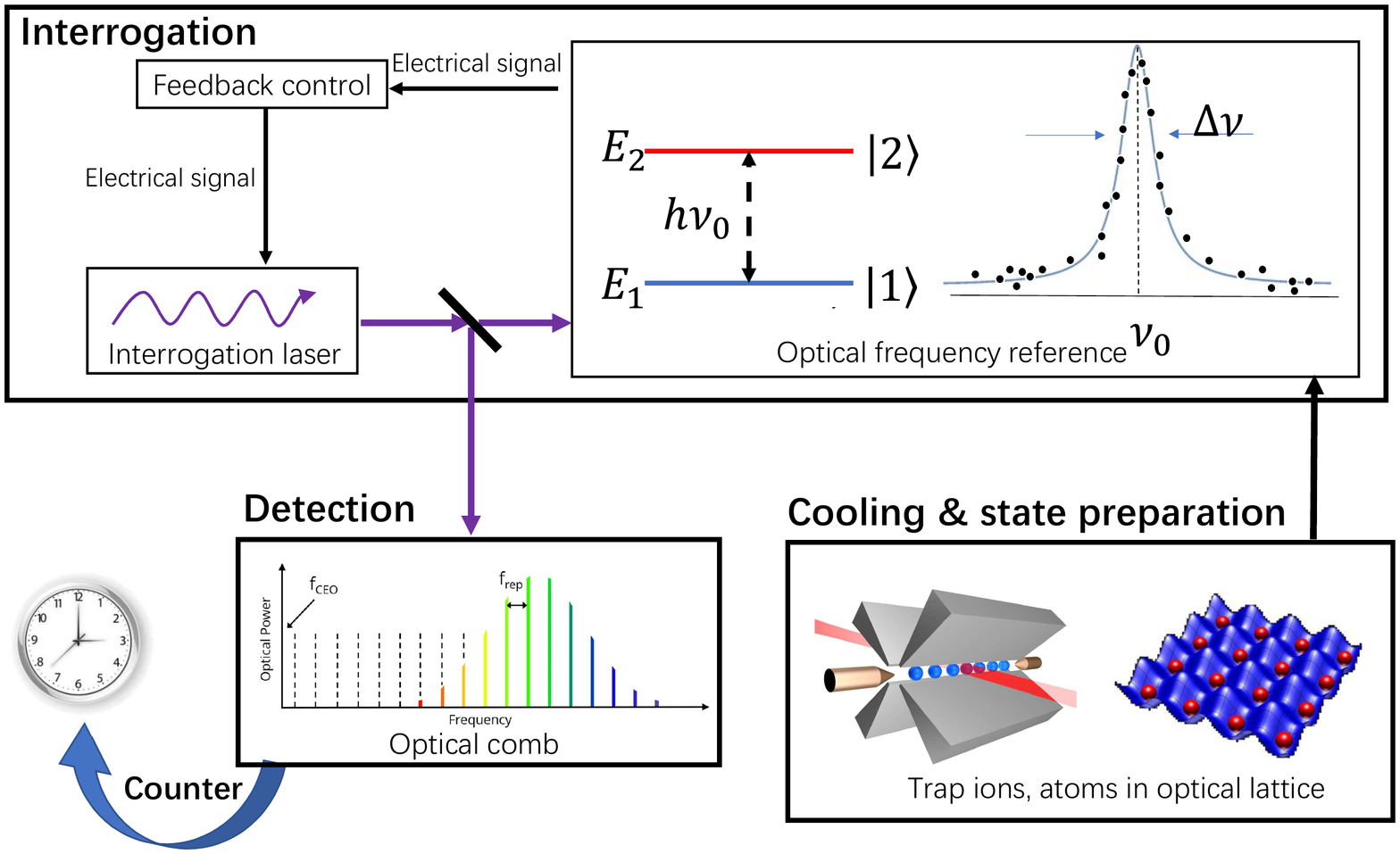}
\caption{Schematic diagram of an optical atomic clock~\cite{poli2013optical}. Atomic spectroscopy is measured from the optical lattice trapped atoms or the trap ion. The interrogation laser is precisely locked to the atomic transition. An error signal is derived from atomic spectroscopy that is fed back to the laser for closed-loop locking. An optical frequency comb, as a counter, converts the optical frequency to a microwave that can be used as the standard frequency of time.}
\label{optical clock}
\end{figure*}

	An optical atomic clock conceptually resembles a classical pendulum clock. In a pendulum clock, a pendulum swings periodically with small amplitude as a local oscillator, the wheels and clock hands together form a counter, and a watchmaker with good a clock correcting the pendulum clock regularly is considered to be a stable reference. In an optical atomic clock, the interrogation laser generates periodically oscillating electromagnetic wave to function as a local oscillator, the frequency is measured by the optical frequency comb as a counter, and trapped neutral atoms or ions take the role of a stable reference by interacting and correcting the frequency of interrogation laser through laser spectroscopy. That is, as shown in Fig.~\ref{optical clock}, one cycle of optical frequency standards includes the following three steps: (1) cooling and state preparation, (2) interrogation, and (3) detection and signal processing. Atoms and ions can be cooled and prepared in an optical lattice or ion trap, respectively. Then, an interrogation laser will be locked to the optical transition {from extracting} the optical frequency through the given optical transitions. In the process of detection and signal processing, an optical frequency comb is used as a counter to obtain the laser frequency precisely. This process converts the optical frequency into countable microwave or radio frequency signals as a frequency standard for many practical applications.

	There are two parameters to show the performance of the clock: stability and uncertainty. Stability is the fluctuation of the standard frequency of the clock over a long period of time, which is determined by the physical system and measurement in nature. It is generally described by the Allan variance defined as~\cite{ludlow2015optical}
	\begin{equation}
	\sigma_{y}^{2}(\tau)=\frac{1}{2(M-1)} \sum_{i=1}^{M-1}\left[\langle y(\tau)\rangle_{i+1}-\langle y(\tau)\rangle_{i}\right]^{2},
	\end{equation}
	{where $\langle y(\tau)\rangle_{i}=\langle {\Delta \nu(\tau) }/{\nu_{0}}\rangle_{i}$ is the $i^{\rm th}$ measurement of the average fractional frequency difference over duration $\tau$, $\nu_{0}$ is the frequency of the reference transition, and $\Delta \nu$ is the frequency error.} Notice that a higher stability of an optical clock corresponds to a smaller value of Allan variance. Stability can also be described as~\cite{poli2013optical}
	\begin{equation}
	\sigma_{y}(\tau) \approx \frac{\Delta \nu}{\nu_{0} \sqrt{N}} \sqrt{\frac{T_{c}}{\tau}},
	\end{equation}
	where $\Delta \nu$ is the spectral linewidth of the clock system, $N$ is the number of atoms or ions used in a single measurement, $T_{c}$ is the time required for a single measurement cycle, and $\tau$ is the total measurement duration. In the expression above, ${\nu_{0} }/{\Delta \nu}$ can be understood as a quality factor $Q$, ${1}/{\sqrt{N}}$ is the atomic measurement projection noise which is also referred as the standard quantum limit (SQL), and ${\tau}/{T_{c}}$ is the number of successive measurements. 
	The biggest advantage of an optical clock is that it has a higher $Q$ value, which is in general 5 orders of magnitude higher than a microwave clock. For instance, a Cs atomic clock that defines the Second has a $Q$ value of only $10^{10}$, with transition frequency of \SI{9.19}{\giga\hertz} and linewidth of \SI{1}{\hertz}. As a comparison, the $^{88}$Sr$^+$ optical clock choosing a 674nm quadrupole transition has a linewidth of \SI{0.4}{\hertz} and a quality factor up to $10^{15} $. By far, the $^{27}$Al$^+$ quantum logic clock is the most precise clock in the world with $Q$ of ${1.4 \times 10^{17}}$~\cite{brewer2019al+} and a narrow linewidth of {\SI{8}{\milli\hertz}}. Besides, the \SI{467}{\nano\metre} electric octuple transition between $^2$S$_{1/2}$ and $^2$F$_{7/2}$ levels of $^{171}$Yb$^+$ ions features a natural linewidth in the nanohertz range and a quality factor of $10^{23}$.
	
	Another important parameter of an optical clock is frequency accuracy, {which is a systematic error due to the perturbation of the undesired fields.} There are many factors that affect the frequency instability, which can be roughly divided into the frequency shift caused by external environment (e.g., electric field, magnetic field and blackbody radiation) and the relativistic effect (gravitational shift)~\cite{ludlow2015optical}.
	The stray electric field and magnetic field felt by the atoms are the main factors that cause the energy level shift. Owing to the Zeeman effect, the clock transition frequency will be affected by fluctuation of a magnetic field. The frequency shift $\Delta \nu$ can be expressed as
	\begin{equation}
	\Delta \nu=\nu-\nu_{0}=C_{ 1} B+C_{ 2} B^{2}+\cdots.
	\end{equation}
	A common method to suppress the first-order Zeeman shift is to alternately interrogate two symmetrically shifted Zeeman components. For the second-order Zeeman shift, one has to stabilize the magnetic field as much as possible. For example, in a $^{87}$Sr lattice clock, Bloom {\it et al.}~\cite{bloom2014optical} modulate the clock transition and extract from it an error signal for the stabilization of the magnetic field, and successfully reduce the frequency shift below $10^{-18}$.
	The AC stark shift induced by lasers is another source of fluctuation. For neutral atoms trapped by laser fields, the wavelength can be selected specifically so that the polarizabilities of the ground state and the excited state are the same, thereby eliminating the AC Stark shift. This special wavelength is called magic wavelength. For ions, the AC stark shift induced by the interaction between laser light and ions must be accounted for. Another type of Stark shift is the blackbody radiation (BBR) shift. The electric field associated with the thermal radiation emitted by the trap structure around atoms and ions can cause a quadratic Stark shift. For example, the BBR shift of the $^5$S-$^4$D$_{5/2}$ clock transition in $^{88}$Sr$^+$ is calculated to be \SI{0.250(9)}{\hertz} at room temperature ($T=$\SI{300}{\kelvin}) and the relative frequency shift is 5.6$\times$10$^{-16}$~\cite{jiang2009blackbody}. Thus, if one wants to reach a frequency uncertainty as low as 10$^{-18}$, the BBR shift must be considered. Similarly, gravitational shift limits the increase in frequency accuracy. For example, when comparing frequencies from two optical clocks, a height difference of $\Delta h =$ \SI{10}{\centi\metre} can result in a frequency shift of $\delta f/f_0=10^{-17} $.
	
	\subsection{Recent progresses}
	Up to now, the most accurate optical clocks we have ever built are based on neutral atoms in optical lattices and single trapped ion~\cite{ludlow2015optical,poli2013optical}, owing to the large signal-to-noise ratio and extremely well isolation from external environment. Thus, in this section we mainly focus on progresses on these two types of clocks, together with some brief introduction about the transportable clocks which are of particular interest in many application scenarios. Fast and exciting developments are also witnessed in other types of optical clocks, such as optical clocks based with expanding atoms~\cite{sengstock1994optical,friebe2008absolute,wilpers2006improved,degenhardt2005calcium, ido2005precision}, active optical clocks with line width as narrow as MHz~\cite{chen2009active,norcia2018frequency,zhuang2015research,zhang2013active,zhuang2014active}, highly charged ion optical clock that is not sensitive to environmental impact~\cite{yu2018selected, derevianko2012highly,safronova2014highly,dzuba2012high,kozlov2018highly}, nuclear optical clock with high clock transition frequency~\cite{campbell2012single,thielking2018laser}, and ion optical clock with multiple ions~\cite{arnold2015prospects,herschbach2012linear,keller2016evaluation,schulte2016quantum}, but are not included in the present paper.
	
	\subsubsection{Optical lattice clock}
	\begin{figure}[t]
		\centering
		\includegraphics[width=9cm]{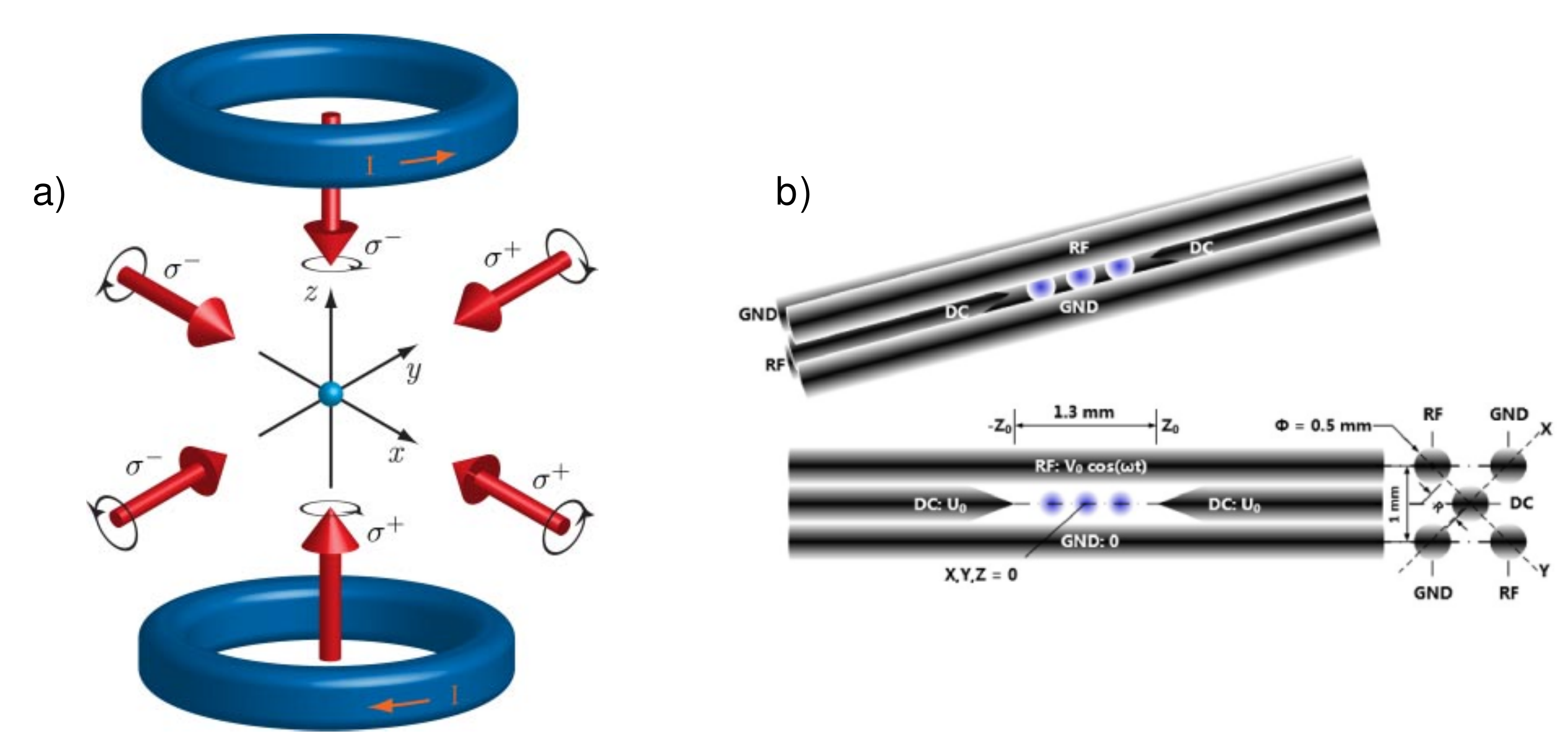}
		\caption{a) Sketch of the Magneto-optical trap (MOT) setup~\cite{poli2013optical}. Three pairs of retro-reflected laser beams cross each other at the center of the trap. A pair of anti-Helmholtz coils provide the necessary quadrupole magnetic field for trapping. The atomic cloud is collected in the center of the trap. b)The diagram of one linear 4-rod trap(adapted from Ref.~\cite{xiangzhang2015quantum}).}
		\label{fig2}
	\end{figure}
	One of the most common optical atomic clocks is the optical lattice clock. Optical lattice {in this clock} is used for several reasons. First, it decouples the external and internal degrees of freedom of the atoms, so there is no dominating effect such as Doppler shift of atomic motion to the clock transition. The high thermal velocity of atoms at room temperature can cause a Doppler shift of \SI{1}{\giga\hertz} or even higher, which is 12 orders of magnitude greater than the milli-hertz uncertainty expected by the community. Second, it avoids the interaction shift of optical transition due to collisions in a dense atomic cloud. In addition, it also provides an effective isolation of the atoms from the outside environment. Before loading atoms into an optical lattice, one has to use a standard laser cooling techniques, such as magneto-optical trap (MOT) to collect atoms and cool the temperature before they can be trapped. As illustrated in Fig.~\ref {fig2}(a), a MOT restricts atoms through a combination of laser beams and magnetic field gradient. In the MOT, three pairs of orthogonal counter-propagating laser beams are red-detuned to a strong cyclic transition. {The laser beams in the ``capture zone"} provide a viscosity force in the opposite directions of the atomic motion. A pair of anti-Helmholtz coils provide a magnetic field gradient and cause the spatial dependence of the scattering force exerted on atoms, towards the center of the trap. In this configuration, millions of atoms are collected in a second, and the temperature is cooled down to less than \SI{1}{\milli\kelvin}. However, the MOT beams and magnetic field can significantly shift the clock transition. When the atoms are loaded into an optical trap based on red-detuned standing waves, which form a periodic lattice potential for the atoms, it is easy to reach the Lamb-Dicke region where the atomic motion must be treated quantum mechanically. In an optical lattice with red-detuned lasers, atoms are attracted to the high-intensity region of the lattice light and oscillate around the anti-nodes of the standing wave. 
	In 2003, H. Katori {\it et al.} proposed to use $^{87}$Sr atoms trapped in an optical lattice to build an ultra-stable optical lattice clock~\cite{katori2003ultrastable}. After 10 years of developing, an optical atomic clock with strontium atoms finally showed better performance than Cs atomic clocks in 2013~\cite{le2013experimental}. This new clock reached a total uncertainty of $1.5 \times 10^{-16} $.

	By tuning the lattice light to the magic wavelength, the frequency shift of clock transition due to the lattice can be minimized to produce an ultra-narrow, minimally perturbed spectrum. In the same paper by H. Katori~\cite{katori2003ultrastable}, the magic wavelength optical lattice trapping technology was firstly proposed. The essence of this design is to construct an optical lattice with a carefully chosen laser frequency, called magic wavelength, which can produce the same AC Stark shift to the upper and lower energy levels of the clock transition. Thus, the Doppler and recoil frequency shifts can be eliminated without introducing additional frequency shifts. The first magic wavelength lattice strontium atomic clock was successfully demonstrated in 2005~\cite{takamoto2005optical}. In 2018, the Katori group proposed a concept of magic light intensity in which the total light shift is canceled about 30$\%$ of a lattice-intensity variation~\cite{ushijima2018operational}. By obtaining electric-quadrupole and magnetic-dipole polarizabilities difference through the experiment, one derives two distinctive operational conditions that make the total light shift insensitive to lattice intensity variation at the $10^{-19}$ level. 
	
	With the ``magic" of magic wavelength, optical lattice clock has been realized in many atomic species and the stability has stepped into the region of $10^{-18}$. In 2013, an ytterbium optical lattice clock developed by the Ludlow group at NIST achieved an unprecedented atomic clock stability of $1.6\times 10^{-18}$ after 7 hours of averaging~\cite{hinkley2013atomic}. In 2018, the stability was further improved to $3.2 \times 10^{-19}$, with a system uncertainty of $1.4 \times 10^{-18}$~\cite{mcgrew2018atomic}. In 2014, Jun Ye's group at NIST demonstrated a strontium optical lattice clock with stability and uncertainty of the order of $10^{-18}$~\cite{bloom2014optical,nicholson2015systematic}. In 2017, they developed a fermionic strontium optical lattice clock in a three-dimensional optical lattice~\cite{campbell2017fermi}, and greatly reduced density-dependent frequency shifts to achieve a measurement precision of $5\times  10^{-19}$ in one hour of averaging time. In 2019, they compared the one-dimensional optical lattice clock with the three-dimensional strontium optical lattice clock~\cite{oelker2019optical}, and the stability of $4.8\times 10^{-17}/\sqrt{\tau}$. The measurement precision reaches $6.6\times 10^{-19}$ after an hour average time.
		
	At the same time, scientists in China also contributed a great deal to the development of optical atomic clocks. In 2015, the National Institute of Metrology of China completed the closed-loop locking of a strontium atomic optical clock, and achieve a self-comparison stability of $6.6\times 10^{-15}/\sqrt{\tau}$ and a system uncertainty of $2.3\times10^{-16}$. The measurement of the absolute frequency was adopted by the Certificate in Investment Performance Measurement (CIPM)~\cite{yi2015first}. In 2015, the National Time Service Center of Chinese Academy of Sciences successfully prepared a cold atomic sample of bosonic $^{88}$Sr and realized a magnetic induction~\cite{lu2017exploration} detection of the clock transition line~\cite{xu2015observation}. Two years later, the closed-loop operation of $^{87}$Sr optical lattice clock achieved a stability of $5\times 10^{-15}/\sqrt{\tau}$ and reached $5.7\times 10^{-17}$~\cite{wang2018strontium} after integration time of 3000s. Through further improvement of the system, continuous stable closed-loop operation was achieved for more than 8 hours in 2018. The self-comparison measurement showed a stability of $ 1.6\times 10^{-15}/\sqrt{\tau}$, and $2.8\times 10^{-17}$ in 2000s. In 2016, the East China Normal University completed the closed-loop locking of an ytterbium atomic optical clock with stability of $2.9\times 10^{-15}/\sqrt{\tau}$ and a system uncertainty of $1.7\times 10^{-16}$~\cite{gao2018systematic}. In 2017, the ytterbium atomic optical clock at the Wuhan Institute of Physics and Mathematics of Chinese Academy of Sciences achieved closed-loop locking with a stability of $ 2.4\times 10^{-14}/\sqrt{\tau}$~\cite{liu2017realization}.

	\subsubsection{Trapped ion optical clock}
	
	A trapped-ion system uses an oscillating (radio frequency) electric fields to confine ions at their dynamic equilibrium positions after the ions are laser cooled. This technique was put forward by Wolfgang Paul and Hans Dehmelt~\cite{paul1990electromagnetic,major1968exchange}, for which they won the Nobel prize of 1989. As illustrated in Fig.~\ref {fig2}(b), a 4-rod trap, so called Paul trap, uses four electrodes to form a rotating radio frequency electric field. The potential of this field can be described as a parabolic pseudo-potential on the radial ($x$-$y$) plane, where the ions are elastically bound to the axial $z$-axis. In the $z$-direction, another two electrodes generate a static coulomb potential, so that ions could be arranged into a string. 
	
	The ions can be cooled via Doppler cooling technique with a red-detuned laser beam. However, the lowest temperature attainable of such a process is the Doppler limit, which is generally less than \SI{1}{\milli\kelvin}. Sideband cooling can further cool ions to the ground state of motion~\cite{ludlow2015optical}. At Doppler limit, the range of ionic motion are much smaller than the wavelength of the probe light, such that the ionic motion must be treated quantum mechanically as quantum harmonic oscillators. In this so-called ``Lamb-Dicke" regime, the state of an ion can be written as $\ket{\uparrow(\downarrow),n}$, where $\ket{\uparrow}$ and $\ket{\downarrow}$ are internal levels, and $\ket{n}$ labels the Fock states of phonon. Based on the coupling between phonon and internal levels, the red-sideband transition $\ket{\downarrow,n} \rightarrow \ket{\uparrow,n-1}$ decreases the phonon number by 1. If one applies an optical pumping beam after the red-sideband transition, all population of spin-up states $\ket{\uparrow,n-1}$ is flipped to their corresponding spin-down states $\ket{\downarrow,n-1}$. As a result, the average number of phonon decreases by 1. One can repeat this process many times until the ion is cooled down to the ground state of motion. Except this sideband cooling mechanism, the electromagnetically-induced-transparency (EIT) cooling provides another efficient way to reach for the quantum ground state of ions~\cite{feng2020efficient}.
	
	The laser cooling technique is only applicable for ions of suitable energy level structure, such as Be$^+$, Mg$^+$, Ca$^+$, Cd$^+$ and Ba$^+$. For other types of ion without such desirable energy levels, one can rely on sympathetic cooling. The idea of sympathetic cooling is to trap the target ions together with an auxiliary component which can be laser cooled, so that the thermalization of the auxiliary medium with the target ion will bring it to a low temperature. In this process, the auxiliary component, usually called the cooling ion, is maintained at low temperature by laser cooling. The target ion, called the cooperative cooling ion, transfers its own momentum through Coulomb interaction to the cooling ion to achieve the cooling effect. In 2005, Schiller group demonstrated sympathetic cooling of He$^+$ with the aid of Be$^+$~\cite{roth2005sympathetic}. Another example of the application of sympathetic cooling in trapped ion clock is the Al$^+$ clock, in which case the cyclic transition between internal levels is in the deep ultraviolet regime. In 2019, NIST demonstrated a successful sympathetic cooling and state readout in an Al$^+$ ion trap by using Mg$^+$ as auxiliary ions~\cite{brewer2019al+}. In this work, they eliminated the heating of long-term motion with a new trap. They also implemented Doppler cooling and sideband cooling of Al$^+$ and Mg$^+$ ions within \SI{14}{\milli\second} to prepare them in the three-dimensional ground state. At the same time, the work also considered the additional micro-motion, time dilate shift, blackbody radiation shift and second-order Zeeman effect. Through the above efforts, the systematic uncertainty of the clock has firstly reached an unprecedented $9.4 \times 10^{-19} $ and frequency stability $1.2 \times 10^{-15} /\sqrt {\tau} $.
	
	Because the cooling process takes as long as \SI{14}{\milli\second}, the detected duty cycle in NIST experiment is not efficient enough. Recently, the Monroe group used EIT cooling to sub-Doppler cool a large number of ion chains (including about 40 ions) to the three-dimensional motion ground state within \SI{300}{\micro\second}~\cite{feng2020efficient}, which improves the cooling speed by roughly a factor of 5. This may shorten the detection time and further reduce the systematic uncertainty.
	
	Currently, there exist many optical clocks with different trapped-ion worldwide, including Al$^+$, $^{40}$Ca$^+$, $^{88}$Sr$^+$, $^{171}$Yb$^+$, $^{199}$Hg$^+$, $^{138}$Ba$^+$$^{115}$In$^+$~\cite{ludlow2015optical}, and the efforts to push the limit of trapped-ion optical clocks will not stop.

	\subsubsection{Transportable clock}
	
	The applications of optical clocks in geodetic and space navigation set a stringent demand for their transportability~\cite{delehaye2018single}. To develop a transportable or space optical clock, one has to redesign or modify some key components for the requirement of compactness and reliability, while not to compromise too much in its performance. This usually means a drastic reform of the entire laser, vacuum, and ultra-stable chamber systems. 
	
	For ion optical clocks, Cao {\it et al.}~\cite{cao2017compact} re-engineered the $^{40}$Ca$^+$ optical clock into two subsystems: a compact single ion unit and a compact laser unit,  and realized an optical clock of \SI{0.54}{\metre}$^3$ in volume. The system fractional uncertainty was estimated at $ 7.8 \times 10^{-17}$. In 2020, the Wuhan Institute of Physics and Mathematics suggested a method of integrating multiple wavelength stabilization on a multi-channel cavity and provided a scheme for compact laser units in transportable optical clocks~\cite{wang2020integrated}.
	
	Regarding the optical lattice clock, the LENS group built a transportable $^{88}$Sr light clock in a volume less than \SI{2}{\metre}$^3$ in 2014, and demonstrated a frequency uncertainty of $ 7.0 \times 10^{-15}$~\cite{poli2014transportable}. In 2017, PTB installed a $^{87}$Sr optical clock in an air conditioned car trailer. Its systematic uncertainty is $7.4 \times 10^ {-17}$ against a stationary lattice clock, and an instability of $ 1.3\times 10^{-15}/\sqrt{\tau}$ with an averaging time $\tau$ in seconds~\cite{koller2017transportable}.
	
	The development of integrated optical comb is also in progress, and technologies such as micro-cavity optical comb are also developing rapidly. At present, a compact optical-clock architecture is proposed with significant reduction both in component size and complexity by the integration of silicon-chip photonics~\cite{newman2019architecture}. Meanwhile, in order to further reduce the size of the clocks, it is also necessary to miniaturize the remaining parts of the clocks, including controlling electronics and optical components. With new digital electronic technology, control electronics can now become extremely compact. Recently, more efforts have been focusing on minimizing the optical frequency combs~\cite{kippenberg2011microresonator,lezius2016space}, laser sources~\cite{burd2016vecsel,barwood2012automatic,liang2015ultralow} and integrated optics~\cite{eltony2013transparent,mehta2016integrated} for both optical lattice and ion trap clocks.
	
	\subsubsection{Heisenberg limit}
	
	In all types of clocks, the measurement is bounded by the standard quantum limit (SQL) $1/\sqrt{N}$ because of the quantum projection noise~\cite{itano1993quantum}. One then has to step into the regime of quantum metrology to reach the Heisenberg limit $1/N$, which is based on the Heisenberg uncertainty relation~\cite{ludlow2015optical}. There are usually two ways to reduce the frequency uncertainty to purse the Heisenberg limit:
	\begin{enumerate}
		\item Reduce the molecular uncertainty by preparing squeezed states~\cite{kitagawa1993squeezed};
		\item Increase the spin response to frequency by utilizing the largest entangled state, or the Greenberger-Horne-Zeilinger (GHZ) state $\psi_{\mathrm{GHZ}}=\left(\left|\downarrow_{1} \downarrow_{2} \ldots \downarrow_{N}\right\rangle+ |\uparrow_{1} \uparrow_{2} \ldots \uparrow_{N}\rangle\right)/\sqrt{2}$.
	\end{enumerate}
	Along the second route of preparing entangled states in trapped ion systems, the Kim group developed a definite method to generate arbitrary phonon NOON states~\cite {zhang2018noon}, and experimentally realized 9 phonon NOON states in a single ion in 2016. Moreover, it is observed that as $N$ increases, the lower bound of the Heisenberg limit is reached. By increasing the number of ions, the Kim group implemented a scalable global entanglement gate, and prepared the GHZ state in an entanglement operation in 2019~\cite{lu2019global}. They successfully demonstrated a multi-partite entanglement in a system of up to 4 qubits, and reached a state fidelity of of $93.4\%$. These realization of entangled states pave the way to approach the Heisenberg limit, which may further improve the performance of the ion optical clock.
	
	\section{Magnetic field measurement}
	\label{sec:Bfield}
	
	Magnetic field is one of the most common but also fundamental physical quantities in all electromagnetic events in nature. Measurement of a magnetic field is widely used technique in many applications including aerospace and deep-sea exploration, mineral discovery, earthquake monitoring, biology, and biomedical science~\cite{budker2007optical, kimball2013optical}. A conventional magnetic sensor employs the classical electromagnetic coupling between the field and the probe, and can realize a measurement of magnetic field within a wide dynamic range via a relatively simple mechanical and electric structure. For example, fluxgate magnetometer and Hall-effect sensor have a detection range from \SI{1}{\nano\tesla} to \SI{e7}{\nano\tesla}. The sensitivity, however, is usually limited to \SI{1}{\nano\tesla}/$\sqrt{\rm Hz}$. In comparison, quantum magnetic sensors are developed by quantum technologies, which have very high sensitivities when measuring a weak field. For example, optically pumped magnetometer, nuclear-precession magnetometer and Overhauser magnetometer can measure the magnetic field from \SI{0.1}{\nano\tesla} to \SI{e5}{\nano\tesla}. One of the most successfully developed equipment for this measurement is low temperature superconducting quantum interference device (SQUID) magnetometer, which has reached a sensitivity level of \SI{1}{\femto\tesla}/$\sqrt{\rm Hz}$. However, the requirements of complicated and expensive cryogenic system, and low spatial resolution of SQUID magnetometer significantly limit its applications. In the past 20 years, atomic magnetometers based on neutral atoms has demonstrated compact, non-cryogenic alternative to sub-femtotesla-sensitity measurement of weak magnetic field.
	
	In this section, we focus on the magnetic field measurement based on cold atoms. The word ``cold atoms" is used for all quantum sensors based on either cold atomic ensembles with reduced distribution of thermal velocity~\cite{hosten_measurement_2016,bohnet_reduced_2014}, or Bose-Einstein condensates(BEC) as a macroscopic quantum state~\cite{kruse_improvement_2016,muessel_twist-and-turn_2015}.
	
	\subsection{Principles}
	Over 170 years ago, M. Faraday noticed that the polarization of a linear polarized light was rotated when it went through a medium in a magnetic field along the propagation direction~\cite{faraday1846xlix}. This phenomenon, known as Faraday rotation, is attributed to the different refractive index of different circular polarization component of the linear polarized light in the medium. This then leads to different accumulated phase shifts and consequent rotated polarization of the combined beam with linear polarization 
	\begin{equation}
	\phi=2\pi\frac{(n_{+}-n_{-})z}{\lambda}, 
	\end{equation}
	where $n_{\pm}$ are refractive indices of different circular polarization components, $z$ is the propagation distance, and $\lambda$ is the wavelength of the laser. This effect thus provides a mechanism to detect the intensity of a magnetic field if a proper medium is identified to have a strong Faraday rotation. 
	
	Microscopically, the mechanism of atomic magnetometer relies on Larmor precession, which is the precession of the magnetic momentum of a particle about an external field. The strength of the magnetic field determines the precession frequency, called the Larmor frequency. With the technologies of laser spectroscopy, the Larmor frequency could be measured in spectroscopy. This basic principle is a quite universal mechanism for most of the atomic magnetometers, no matter optical pump technique is utilized of not. 
	
	Facilitated by the fast development in technology to control atoms with laser light, atoms with population polarization was used to detect magnetic field. The first experiment was demonstrated in ${^4}$He in the 1960s~\cite{keyser1961metastable}. Since then, optical pumped ${^4}$He magnetometer has been widely used over tens of years. In 2003, a new spin-exchange-relaxation-free(SERF) atomic magnetometer with Potassium atoms at room temperature vapor cell was developed by Romalis~\cite{kominis2003subfemtotesla}. In this new setup, an unprecedented magnetic field sensitivity of \SI{0.54}{\femto\tesla}/$\sqrt{\rm Hz}$ was achieved, and a 2mm spatial resolution was obtained. This performance shows an alternative device than SQUID magnetometer for weak field detection, but in a much simpler way. In 2012, a sensitivity of \SI{20}{\femto\tesla}/$\sqrt{\rm Hz}$ was obtained in an atomic magnetometer based on a micro-fabricated vapor cell of $^{87}$Rb atoms, and a micro-machined silicon sensor head was demonstrated~\cite{mhaskar2012low}. Now many variants of optical pump atomic magnetometers have been realized in labs, and started to be applied in geophysics and biomedicine.
	
	\subsection{Recent progresses on cold atom magnetometers}
	Cold atoms are suggested to be an alternative way to further improve the performance of atomic magnetometer, owing to the fine control of various fluctuation sources such as negligible Doppler broadening of the optical transition, significantly smaller diffusion, and much longer coherence time of atoms. Besides, the strong suppression of collisions, high local density and higher spatial resolution also make the system an interesting system to test new quantum technologies for further improving the performance as atomic magnetometers~\cite{kimball2013optical}.
	
	\subsubsection{Trapped non-degenerate gases}
	The earliest experiment of magnetic field measurement with cold atoms was performed in \num{9e7} laser cooled $^{87}$Rb atoms\cite{isayama1999observation}, in which Larmor spin precession was observed through paramagnetic Faraday rotation. In this experiment, a pump beam polarized the spin population and the polarization of a probe beam was monitored to detect the Larmor frequency, so that the magnetic field was determined. The temperature of the atoms was \SI{10}{\micro\kelvin}, the traditional Doppler broadening and collisional broadening with buffer gas were absent, therefore a high signal to noise ratio was obtained and a precision of \SI{18}{\pico\tesla} was reported. This is the first demonstration that laser cooled atoms could be used to improve the performance of magnetic field measurement. Since then, atomic magnetometers using non-degenerate cooled gases have been realized in different traps and a sensitivity down to {\SI{10}{\pico\tesla}/$\sqrt{\rm Hz}$} and a spatial resolution to \SI{50}{\micro\metre} were reported~\cite{terraciano2008faraday, fatemi2010spatially, koschorreck2010sub,koschorreck2011high, deb2013dispersive, eliasson2019spatially}.
	
	\subsubsection{BEC in atomic chip}
	Since 2006, $^{87}$Rb Bose-Einstein condensates on an atom chip were demonstrated to be sensitive sensors for magnetic and electric fields. With {\it in situ} imaging technique, the local density of BECs was obtained with high spatial resolution, such that the two-dimensional magnetic field distribution above the micro-structured atomic chip was extracted~\cite{wildermuth2006sensing, wildermuth2005microscopic, aigner2008long}. In these setups, both high sensitivity of \SI{1e-10}{\tesla} and high spatial resolution of about $1\mu$m have been reached, providing new possibilities of simultaneous observation of microscopic and macroscopic phenomena.
	
	\subsubsection{Spinor BEC}
	In 2007, the Stamper-Kurn group demonstrated a precise measurement of the Larmor precession of in a $^{87}$Rb spinor Bose-Einstein condensate~\cite{vengalattore2007high}. In this work, the Larmor precession was induced by a radio-frequency (RF) pulse. The Larmor precession phase was obtained with magnetization-sensitive phase contrast imaging of the spinor condensate, in which a detuned circular polarized imaging light is used. In this cold atom magnetometer, the best field sensitivity obtained was \SI{8.3}{\pico\tesla}/$\sqrt{\rm Hz}$ at a {spatial resolution of \SI{120}{\micro\metre\square}}.
	
	\subsubsection{Spin echo}
	In 2013, Y. Eto demonstrated an atomic magnetometer with the spin-echo technique in $^{87}$Rb $F = 2$ Bose-Einstein condensates. In this experiment,  the RF Hahn-echo pulse sequence ($\pi/2$-$\pi$-$\pi/2$) was used to rotate the spin vector and implement the spin echo, so that the effect of undesirable inhomogeneities and stray magnetic fields was minimized. A magnetic field sensitivity of \SI{12}{\pico\tesla}/$\sqrt{\rm Hz}$ of an AC magnetic field was attained at a spatial resolution of \SI{100}{\micro\metre\square}~\cite{eto2013spin}. In a different experiment, the rotary echo-pulse sequence was implemented in a cold gas of Cs atoms~\cite{smith2011three}. Both DC and AC components of the background field along three orthogonal axes were measured, to a resolution of less than {\SI{5}{\nano\tesla}} in a bandwidth of $\sim$\SI{1}{\kilo\hertz}. 
	
	\subsubsection{Radio-frequency magnetometer}
	In 2019, a radio-frequency atomic magnetometer similar to that with thermal vapor cell~\cite{savukov2005tunable} was demonstrated\cite{cohen2019cold}. In this setup, atom clouds of $^{87}$Rb atoms were sub-Doppler laser cooled to \SI{20}{\micro\kelvin} and traditional pump-probe scheme were used. A sensitivity of \SI{330}{\pico\tesla}/$\sqrt{\rm Hz}$ was reported in an unshielded environment. This can potentially provide many applications of RF magnetometer to high spatial resolution regime.
	
	\subsubsection{Spin squeezing}
	In 2012, a squeezing of spin orientation was realized by quantum non-demolition (QND) measurement through a train of short pulses~\cite{sewell2012magnetic}. This led to a spin-aligned atomic ensemble with up to $8.5 \times 10^5$ laser cooled $^{87}$Rb atoms in the $F = 1$ hyperfine ground state and generated spin squeezing and entanglement in the cold atomic ensemble. In this experiment, 3.2~dB of quantum noise reduction and 2.0~dB of spin squeezing were obtained, which improved the short term sensitivity and measurement bandwidth. In 2014, the Oberthaler group demonstrated a scalable spin squeezing through nonlinear dynamics in their Bose Einstein condensates with proper trapping geometries~\cite{muessel2014scalable}. They achieved a suppression of fluctuations by 5.3~dB in 12,300 particles and a single-shot sensitivity of \SI{310}{\pico\tesla}, which corresponds to a sub-shot-noise sensitivity of {\SI{1.86}{\nano\tesla}/$\sqrt{\rm Hz}$}.
	
	\subsubsection{Entanglement}
	Similar to entanglement-assisted magnetometer with thermal vapor cell~\cite{wasilewski2010quantum}, cold atoms magnetometer also borrowed the idea of the entanglement and demonstrated an enhancement of its performance~\cite{ockeloen2013quantum}. In this setup, atomic chip was used to trap and manipulate a small Bose-Einstein condensate of $^{87}$Rb atoms. The entanglement between the atoms was realized by two-body collisions and a state-dependent potential. The measurement overcame the standard quantum limit by 4~dB and obtained an enhanced sensitivity of {\SI{77}{\pico\tesla}/$\sqrt{\rm Hz}$} to microwave magnetic field. In 2020, an interesting 3D magnetic gradiometry was realized in an ultracold atomic scattering halo of pairs in a symmetric entangled spin state~\cite{shin2020entanglement}. A simple magnetic gradiometer was also realized.
	
	\subsubsection{Spatial resolution}
	Compare to the atomic magnetometer based on thermal vapor cell~\cite{budker2007optical}, cold atoms magnetometer have much higher spatial resolution without sacrificing the sensitivity. One example is an early demonstration of high resolution atomic magnetometer with cold atoms~\cite{ koschorreck2011high}, where a sensitivity of {$\SI{10}{\pico\tesla}/\sqrt{\rm Hz}$} at \SI{50}{\micro\metre} spatial resolution was shown. In 2019, a novel design of cold atom magnetometer with spatially-selective and spatially-resolved {\it in situ} measurement was developed~\cite{eliasson2019spatially}. In this new setup, shaped dispersive probe beams and spatially-resolved balanced homodyne detection method were used. These new designs could be used not just to magnetic field sensing, but also to better quantum simulator with quantum gas microscopes.
	
	\subsubsection{Optical lattice}
	Cold atoms trapped in deep optical lattices offer an interesting system for precision measurement of magnetic field. The atoms in the system are well localized with no photon recoil, no collisional effect, so that longer spin coherence time can be attainable. One experiment reported a measurement of Faraday effect of the spin of laser-cooled atoms trapped in an optical lattice~\cite{smith2003faraday}. This may shine new lights on the developing novel atomic magnetometers.
	
	\section{Inertial forces measurement}
	\label{sec:force}
	
	Measurement of gravity related quantities and inertial forces in general has shown many applications from fundamental physics to industry. Due to their ultra-high sensitivity, Raman atom interferometers have been used to measure gravitational acceleration~\cite{kasevich1992measurement}, gravity gradient~\cite{snadden1998measurement}, gravitational constant~\cite{fixler2007atom, lamporesi2008determination}, angular velocity~\cite{riehle1991optical,lenef1997rotation,gustavson1997precision} and they are even used to test general relativity~\cite{fray2004atomic,dimopoulos2007testing}. These devices have great potential in a wide range of circumstances ranging from academic research in fundamental physics, metrology, and geophysics, to industrial applications such as oil and minerals detection, and inertial navigation~\cite{bongs2019taking}.
	
	\subsection{Principles}
\begin{figure}
		\centering
		\includegraphics[width=9cm]{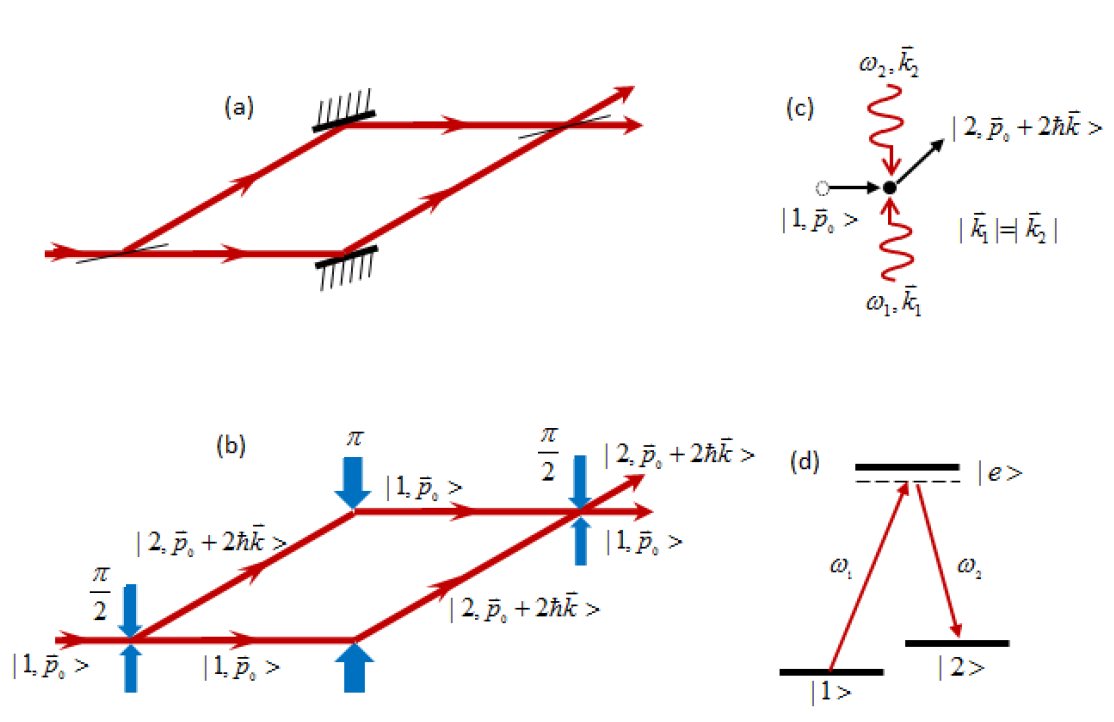}
		\caption{Optical Mach-Zehnder interferometer and Raman atom interferometer. (a) An optical Mach-Zehnder interferometer with beam splitters and mirrors. (b) A Raman atom interferometer with a standard $\pi/2$-$\pi$-$\pi/2$ Raman sequence. (c) Momentum transfer of an atom when its internal state is changed by a Raman pulse. (d) Two-photon Raman process.}
		\label{fig:MZinter} 
	\end{figure}

	It is well known that interference occurs when two light beams overlap with the same frequency, the same polarization and a certain phase difference. The interference pattern changes with varying phase difference between the beams. Similarly, matter waves also exhibit interference phenomena. Raman atom interferometers conceptually resemble an optical Mach-Zehnder interferometer, as shown in Figs.~\ref{fig:MZinter}(a) and \ref{fig:MZinter}(b). In an optical Mach-Zehnder interferometer, the incident light beam is split into two by the first beam splitter. The two beams propagate in different paths, and then redirected by the mirrors to meet and interfere at the second beam splitter. The light power of outputs of the second beam splitter show interference signal, which is used to determine the optical path difference between the two beams. In the Raman atom interferometer, the matter wave of neutral atoms plays the same role as the light beam in an optical interferometer, and the Raman pulses act as beam splitters and mirrors. The coherent splitting, propagation and superposition of the atoms result in the matter wave interference~\cite{peters2001high}, which can serve as an ideal inertial sensor owing to the interaction of atoms and inertial forces.

	Specifically, in a Raman atom interferometer, the two energy levels of an atom are denoted by $\ket{1,\vec{p}_0}$ and $\ket{2,\vec{p}_0+2\hbar\vec{k}}$ as shown in Figs.~\ref{fig:MZinter}(c) and \ref{fig:MZinter}(d). The incident atoms are initially in the $\ket{1,\vec{p}_0}$ state. Being irradiated with a $\pi/2$ (a quarter of a Rabi oscillation) pulsed light, the atoms are transferred to an equally populated superposition of $\ket{1,\vec{p}_0}$ and $\ket{2,\vec{p}_0+2\hbar\vec{k}}$. Atoms in $\ket{2,\vec{p}_0+2\hbar\vec{k}}$ state obtain a momenta $2\hbar\vec{k}$ from the two-photon Raman process, so these atoms are spatially split from the initial $\ket{1,\vec{p}_0}$ state atoms. After a certain time, a $\pi$ pulse (half of a Rabi oscillation) of Raman beams simultaneously interact with all atoms, and reverse the states of the atoms. That is, the $\ket{1,\vec{p}_0}$ state becomes $\ket{2,\vec{p}_0+2\hbar\vec{k}}$ state, while the $\ket{2,\vec{p}_0+2\hbar\vec{k}}$ state atoms goes to $\ket{1,\vec{p}_0}$ state. Once again, after a certain time, the two atom beams simultaneously interact with the second $\pi/2$ pulse, and they both become coherent superposition of $\ket{1,\vec{p}_0}$ and $\ket{2,\vec{p}_0+2\hbar\vec{k}}$ states, so the interference of matter waves occurs. Here, the $\pi/2$ pulse acts as a beam splitter and the $\pi$ pulse as a mirror. The probability that the atoms in one of the states is determined by the differential phase shift of the atoms~\cite{kasevich1992measurement,storey1994feynman}
	\begin{equation}
	P\propto1+\cos{(\Delta\phi_{\rm path}+\Delta\phi_{\rm laser})},
	\end{equation}
	where $\Delta\phi_{\rm path}$ is the phase shift obtained during the propagation in external inertial fields, and $\Delta\phi_{\rm laser}$ is due to the interaction of the atoms with Raman pulses. Through the interaction of atoms with external inertial fields, the phase of the final superposition state of the atoms can determine the inertial fields such as gravitational acceleration. Therefore, the information of gravity field can be extracted from the selective detection of the internal state of atoms.
	
	A practical experimental realization of a Raman atom interferometer includes preparation, manipulation and detection of atoms in certain quantum states. Preparation of atoms usually refers to slowing, cooling and optical pumping of the atoms with laser beams~\cite{metcalf1999laser,foot2005atomic}. These procedures prepare the atoms with small momentum uncertainty in stable ground states for manipulation and detection in subsequent steps. Manipulation of quantum states in a Raman atom interferometer means coherent splitting and recombining the atoms with $\pi/2$ and $\pi$ pulses of two-photon Raman transitions. Once the separated beams propagate in external inertial fields, the phase shifts imprint into the different quantum states of the atoms. State selective detection of atoms reads the information of a certain phase shift by measuring the population of atoms in different quantum states. In the path integral theory, the phase shifts of the atoms interacting with different external fields $\Delta\phi_{\rm path}$ and the Raman pulses $\Delta\phi_{\rm laser}$ can be calculated~\cite{kasevich1992measurement,peters2001high,storey1994feynman, bongs2006high}. The relations between the phase shifts and the external inertial fields can be used to measure the inertial fields.
	
	\textbf{Atomic gyroscope}. Atoms feel the Coriolis force when there is a rotation in the horizontal plane, which results in the Sagnac effect. In this case the phase shifts are
	\begin{eqnarray}
	\Delta\phi_{\rm path} &=& 0,
	\nonumber \\
	\Delta\phi_{\rm laser} &=& 2k_{\rm eff}v\Omega T^{2}+\phi_{1}-2\phi_{2}+\phi_{3},
	\end{eqnarray}
	{where $k_{\rm eff}$ is an effective coupling constant, $\Omega$ is the angular velocity, $v$ is the velocity of the gyroscope, $T$ is the evolution time between two $\pi/2$ pulses}, and $\phi_{1,2,3}$ denote the phase shifts due to Raman pulses.
	
	\textbf{Atomic gravimeter}. When a uniform gravitational field acts on atoms and the direction of Raman light is identical to that of the gravitational acceleration, we have 
	\begin{eqnarray}
	\Delta\phi_{\rm path} &=& 0,
	\nonumber \\
	\Delta\phi_{\rm laser} &=& -k_{\rm eff} g_0 T^{2}+\phi_{1}-2\phi_{2}+\phi_{3},
	\end{eqnarray}
	where $g_0$ is the gravitational acceleration.
	
	\textbf{Atomic gravity gradiometer}. In a gravity field with a uniform gradient when the direction of Raman light is identical to the gravitational acceleration, the phases read
	\begin{eqnarray}
	\Delta\phi_{\rm path} &=& \alpha k_{\rm eff}T^{2} \left( \frac{7}{12}g_{0}T^{2}-v_{0}T-z_{0} \right),
	\nonumber \\
	\Delta\phi_{\rm laser} &=& -k_{\rm eff}g_0 T^{2}+\phi_{1}-2\phi_{2}+\phi_{3},
	\end{eqnarray}
	where $\alpha$ is another coupling constant, and $z_{0}$ is the height variation of the atom trajectory. When the gravitational gradient is measured by an atomic gravity gradiometer close to a well-characterized massive object, the Newtonian gravitational constant can also be precisely determined. The relations above show that the extra inertial fields change the differential phase shift in the Raman atom interferometers, which can be used to measure the inertial fields. 
	
	It is noticed that the differential phase shift of the Raman atom interferometers $\Delta\phi_{\rm total}\propto k_{\rm eff}\propto\lambda^{-1}_{\rm laser}$ if $\phi_{1}-2\phi_{2}+\phi_{3}$ is ignored. That is, the atom interferometer's accuracy is scaled with the laser wavelength, but its sensitivity is much higher than an optical interferometer. Taking the measurement of angular velocity as an example, the additional phase shift is $\Delta\phi=4\pi\Omega \mathbf{A}/\lambda c$ due to the Sagnac effect~\cite{anderson1994sagnac,sagnac1913demonstration,sagnac1913proof}. If we compare the atomic interferometer with the optical interferometer, the differential phase shifts satisfy
	\begin{equation}
	\frac{\Delta\phi_{\rm atom}}{\Delta\phi_{\rm light}}=\frac{4\pi\Omega \mathbf{A}/\lambda_{\rm atom}v}{4\pi\Omega \mathbf{A}/\lambda_{\rm light}c}=\frac{\lambda_{\rm light}c}{\lambda_{\rm atom}v}\sim 10^{11},
	\end{equation}
	assuming the area enclosed by the interferometers $\mathbf{A}$ are the same for both devices. Potentially, the atom interferometer could be much more sensitive than an optical interferometer if they take the same size. Although the potential sensitivity of atom interferometer is currently limited by many technical issues, intensive efforts worldwide have been put to improve the atom interferometer for many applications.
	
	\subsection{Inertial sensors based on atom interferometers}
	
	In the 1950s the idea of the atomic fountain was proposed by Zacharias based on the Ramsey separated fields method~\cite{kasevich1989atomic}, aiming to extend the interaction time between light and atoms~\cite{ramsey1950molecular}. This idea was passed down to generations of physicists by word of mouth but never published because of the unsuccessful attempt in experiments, due to the lack of a high density source of slow atoms~\cite{kasevich1989atomic}. In 1987, the possibility of matter-wave interferometers of low-velocity neutral atoms was proposed, to measure acceleration and rotation as inertial sensors with high sensitivity~\cite{clauser1988ultra}. In the late 1980s, various types of atom interferometers were proposed to be sensitive probes of different physical effects~\cite{borde1983theoretical}, and until the early 1990s the first laboratory demonstration of atom interferometers was realized~\cite{carnal1991young}. Because of their intrinsically high sensitivity to inertial effects, atom interferometers are now widely used as tools for fundamental physics and precision measurements~\cite{cronin2009optics}. 
	
	In 1991, four groups independently realized atom interferometers using different experimental methods. The Mlynek group demonstrated the Young's double-slit experiment of matter waves of He atoms~\cite{carnal1991young}, the Pritchard group used a transmission grating to realize a Na atomic interferometer~\cite{keith1991interferometer}. Bordé {\it et al.}~\cite{riehle1991optical} and the Chu group~\cite{kasevich1991atomic} developed atomic interferometers based on laser beam diffraction of atoms. The former one is referred to Ramsey-Bordé atom interferometer while the later one as Raman atom interferometer. Since the first laboratory demonstrations of atomic interferometers, theoretical and experimental investigation of atomic interferometry developed dramatically~\cite{berman1997atom,cronin2009optics}. 
	In China, studies in this field is initiated by the Zhan group at Wuhan Institute of Physics and Mathematics of the Chinese Academy of Sciences, who realized the first atom interferometer in China in 2005~\cite{ping2007demonstration}. Since then, researches in this field have been very active and many exciting progresses have been achieved. In the following, we will briefly review some important steps for the development of atom interferometer as inertial sensors.
	
	\subsubsection{Atomic gyroscope}
	In 1973, a US patent first came up using a matter wave interferometer to precisely measure a few physics quantities, including the rate of rotation of the apparatus, variation of the gravitational field and magnetic field~\cite{altshuler1976matter}.  
	
	\begin{figure*}
		\centering
		\includegraphics[scale=0.55]{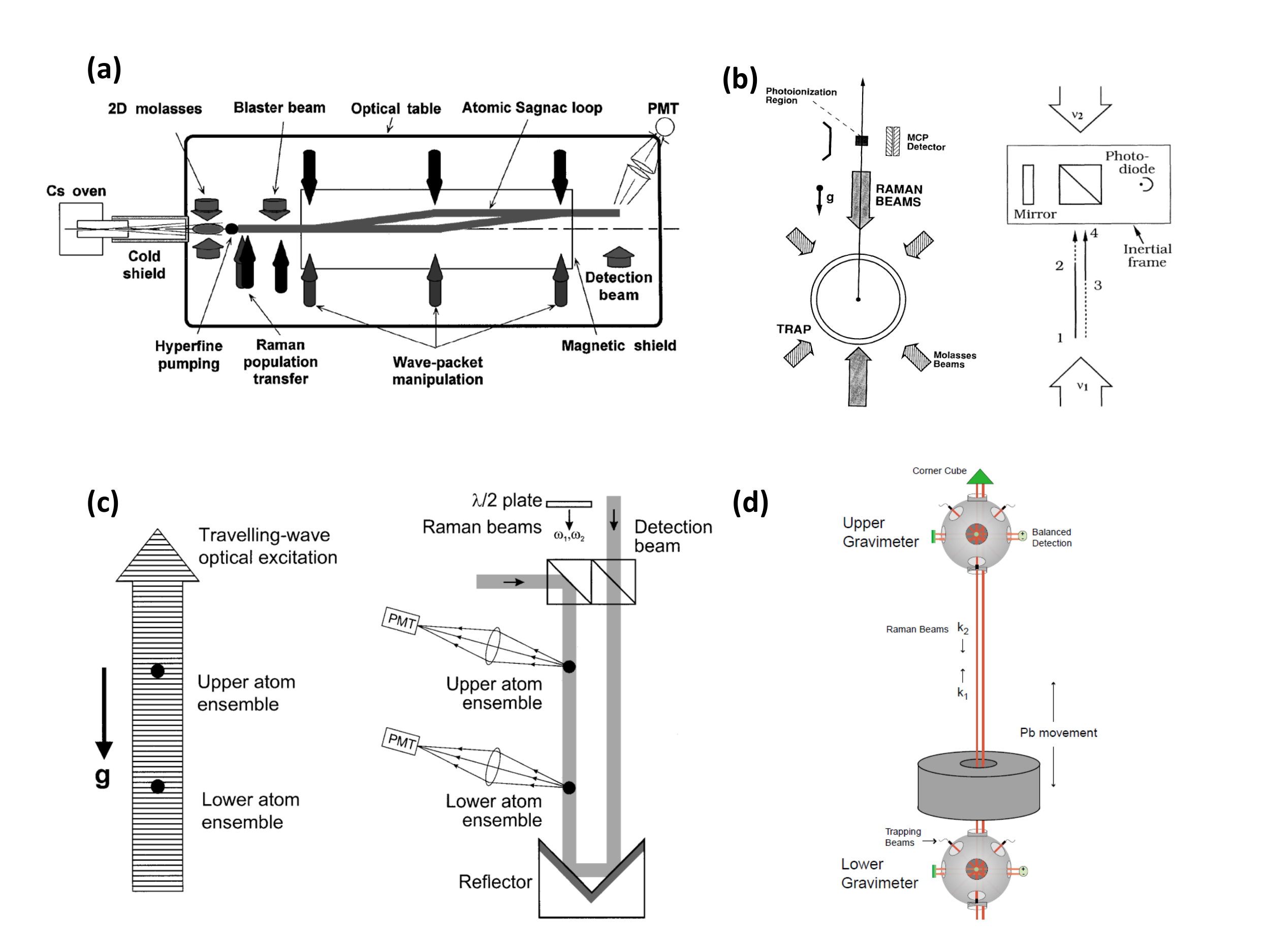}
		\caption{Inertial sensors based on a Raman atom interferometer. (a) An atomic gyroscope for angular velocity~\cite{gustavson1997precision}. (b) An atomic gravimeter for gravity acceleration~\cite{kasevich1991atomic}. (c) An atomic gravity gradiometer for the gradient of gravity acceleration~\cite{snadden1998measurement}. (d) An atomic gravity gradiometer for the Newton gravitational constant~\cite{fixler2007atom}.}
		\label{fig:Ramaninter} 
	\end{figure*}
	
	In 1991, the first Ramsey-Bordé interferometer of a Calcium atomic beam was realized to measure the rotation frequency of the apparatus by F. Riehle {\it et al}~\cite{riehle1991optical}. By rotating their entire apparatus at various rates $\Omega$ and recording the frequency shift of the Ramsey fringes, they demonstrated the first Ramsey atomic gyroscope. In 1997, two groups simultaneously published results of rotation sensing with atom interferometers~\cite{lenef1997rotation, gustavson1997precision}. In Ref.~\cite{lenef1997rotation}, the Pritchard group used a beam of Sodium atoms through nanofabricated transmission gratings. The rotation of an atom interferometer at rates of $-2$ to $+2$ the Earth rates was measured, and $1\%$ agreement with theory was achieved. In contrast, stimulated Raman transitions were used to coherently manipulate atomic wave packets of Cesium atoms in Ref.~\cite{gustavson1997precision} from the Chu group, as shown in Fig.~\ref{fig:Ramaninter}(a). The signal-to-noise ratio of the atomic gyroscope interference fringe was 400:1, and a short-term sensitivity of $2\times 10^{-8}({\rm rad/s})/\sqrt{\rm Hz}$ was more than two orders of magnitude better than the previous results. The Earth's rotation rate was measured with high precision in this setup. Since then this type of Raman atomic gyroscope have been drawing significant attention and widely investigated, because it shows more potential in technical limits, flexibility and sensitivity, compared to the gratings based atomic gyroscopes~\cite{barrett2014sagnac}. 
	
	During 1998 to 2006, the Kasevich group continuously improved the sensitivity of the Raman atomic gyroscope. In 1998, they used two counter-propagating beams of atoms to construct an atomic gyroscope. This new design of dual-interferometer atomic gyroscope took advantage of the same set of Raman beams, which greatly reduced common mode noise and various systematic errors and increased short-term sensitivity to $3\times 10^{-9}({\rm rad/s}) / \sqrt{\rm Hz}$~\cite{gustavson1998dual}. In 2000, they adopted a feedback control system and phase-locked technology in this system, and increased the short-term sensitivity to $6\times 10^{-10}({\rm rad/s}) / \sqrt{\rm Hz}$~\cite{gustavson2000rotation}, which was already a factor of 2 better in sensitivity reported for a \SI{1}{\metre\squared} ring laser gyroscope at that time. In 2006, they fixed the atomic gyroscope directly on the ground of their laboratory with minimal vibration isolation, and used acousto-optic modulators to suppress the spurious phase shift. They demonstrated a gyroscope bias stability of $<$70 $\mu {\rm deg/h}$, scale factor stability of $<$5 ppm and short-term noise $\sim$3 $\mu {\rm deg/h}^{1/2}$~\cite{durfee2006long}, which enabled navigation at a level of system drift much less than 1 km/h. In 2011, the group used a new $\pi/2$-$\pi$-$\pi$-$\pi/2$ Raman pulse sequence in a small and portable setup. This expanded the dynamic range of previous atomic gyroscope by a factor of 1,000. When the angular velocity did not exceed $0.1 {\rm rad/s}$, the gyroscope still maintained a high contrast, and the random drift of the angle remained below 295 $\mu {\rm deg/} \sqrt{\rm h}$~\cite{stockton2011absolute}. This solution simultaneously satisfied the accuracy and the dynamic range, hence make a bridge between lab-based atomic gyroscopes and real devices for inertial navigation and geophysics. There were many other promising progresses for smaller and portable devices, which will be important for future applications~\cite{takase2008precision, muller2009compact, barrett2014sagnac}. 
	Some of the measured performance of atomic gyroscopes reported by different groups are summarized in Tab.~\ref{tab:atomgyro}, including the ones obtained by the Zhan group at Wuhan Institute of Physics and Mathematics of the Chinese Academy of Sciences~\cite{yao2018calibration} and the Hu group at Huazhong University of Science and Technology~\cite{xu2020effects}. 
	\begin{table}
		\centering
		\caption{Measured performance of atomic gyroscopes.}
		\label{tab:atomgyro} 
		\begin{tabular}{|c|c|c|}
			\hline Interrogation time (ms) & Sensitivity $(\mathrm{rad} / \mathrm{s} / \sqrt{\mathrm{Hz}})$ & Reference \\
			\hline 0.085 & $7 \times 10^{-7}$ & {\cite{lenef1997rotation}} \\
			\hline 6.8 & $2 \times 10^{-8}$ & {\cite{gustavson1997precision}} \\
			\hline 6.8 & $6 \times 10^{-10}$ & {\cite{gustavson2000rotation}} \\
			\hline 9.1 & $7.5 \times 10^{-8}$ & {\cite{durfee2006long-term}} \\
			\hline 60 & $2.2 \times 10^{-6}$ & {\cite{canuel2006six-axis}} \\
			\hline 80 & $2.4 \times 10^{-7}$ & {\cite{gauguet2009characterization}} \\
			\hline 4 & $2 \times 10^{-4}$ & {\cite{muller2009compact}} \\
			\hline 260 & $7 \times 10^{-6}$ & {\cite{takase2008precision}} \\
			\hline 206 & $8.5 \times 10^{-8}$ & {\cite{stockton2011absolute}} \\
			\hline 23-25 & $1.2 \times 10^{-7}$ & {\cite{berg2015composite}} \\
			\hline 801 & $3 \times 10^{-8}$ & {\cite{savoie2018interleaved}} \\
			\hline 104 & $1.2 \times 10^{-6}$ & {\cite{yao2018calibration}} \\
			\hline 546 & $1.67 \times 10^{-7}$ & {\cite{xu2020effects}} \\
			\hline
		\end{tabular}
	\end{table}

	\subsubsection{Atomic gravimeter}
	Almost 40 years after Zacharias' original proposal of the atomic fountain, the Chu group at Stanford demonstrated the first working atomic fountain~\cite{kasevich1989rf}, which became one workhorse for next-generation atomic clocks. With the aid of fast developing laser cooling technology, the extension of Ramsey's separated oscillatory field method in the microwave regime to optical domain in the atomic fountain naturally leads to a new design of atom interferometers. Since the accuracy of the Raman atom interferometer is limited by the interacting time of atoms and gravity field, the slow speed of atoms and the long propagation time in the atomic fountain can greatly facilitate the measurement and hence enhance accuracy. 
	In 1991 Kasevich and Chu developed the world's first Raman atom interferometer based on atomic fountain and Raman pulses. They used it to measure the gravity acceleration~\cite{kasevich1991atomic}, as shown in Fig.~\ref{fig:Ramaninter}(b). In this experiment, about $10^{7}$ atoms were initially loaded into magneto-optical trap from slowed Na atom beam, then polarized gradient cooling was used to cool the atoms to about $30\mu$K. The atoms were launched vertically in a moving optical molasses, and optical pumped into the $F = 1$ ground state. The $\pi/2$-$\pi$-$\pi/2$ sequence of Raman pulses was used to manipulate the state of atoms in the fountain. After that, a resonant light ionized the atoms on the $F = 2$ state and the probability of atoms in this state was measured with a micro-channel plate detector. The interference fringe of this atom interferometer was obtained when the frequency of Raman lasers were scanned. From the phase shift caused by the gravity, the gravity acceleration could be determined. They obtained a resolution of $3\times 10^{-6}g$ after 1000 seconds of integration time. Based on this unique design, many groups have been improving this atomic gravimeter for higher resolution. In 1992, Kasevich and Chu obtained $3\times 10^{-8}g$ accuracy after 2000 seconds of integration time, by improving to the Raman scheme in the atom interferometer~\cite{kasevich1992measurement}. In 1999, the Chu group further cooled the atoms, used an active low-frequency vibration isolator and systematically reduced the noise in the atom interferometer. These strategies led to an absolute uncertainty down to $3\times 10^{-9}g$ after 60 seconds of integration time~\cite{peters1999measurement}. In 2008, the Chu group increased the number of Cs atoms to $10^9$ and significantly reduced the atom temperature to 150nK by Raman sideband cooling. The accuracy was increased to $1.3\times 10^{-9}g$ after 75 seconds of integration time~\cite{muller2008atom}. In 2013, the Kasevich group developed a point source interferometer~\cite{dickerson2013multiaxis}. Atom cloud at 3nK, with 30 $\mu$m initial radius was used as an atomic point source, and a spatially resolved detection was implemented by a CCD camera. Combined with a 10m tall atomic fountain, they achieved an unprecedented accuracy of $6.7\times 10^{-12}g$ in one shot of 20 seconds. In China, the development of atomic gravimeter is also fast and exciting. In 2013, the Hu group at Huazhong University of Science and Technology demonstrated an atomic gravimeter and achieved the sensitivity of $4.2\times 10^{-9}g/\sqrt{\rm Hz}$~\cite{hu_demonstration_2013}. In 2018, the same group developed a momentum-resolved detection technique in a sensitive Bragg atom interferometer and improved the resolution for gravity measurements to the level of $7\times 10^{-10}g$ after an integration time of 1000s~\cite{cheng2018momentum}. 
	
	Table~\ref{tab:atomgravi} presents the measured performance of some atomic gravimeters reported by different groups.
	\begin{table}
		\centering
		\caption{Some reported performance of atomic gravimeters.}
		\label{tab:atomgravi}
		\begin{tabular}{|c|c|c|}
			\hline Interrogation time (ms) & Sensitivity ($\Delta g/g/\sqrt{\rm Hz}$)  & Reference \\
			\hline 135 & $9.5 \times 10^{-5}$ & {\cite{kasevich1991atomic}} \\
			\hline 100 & $1.3 \times 10^{-6}$ & {\cite{kasevich1992measurement}} \\
			\hline 320 & $2.3 \times 10^{-8}$ & {\cite{peters2001high, peters1999measurement}} \\
			\hline 100 & $1.4 \times 10^{-8}$ & {\cite{le2008limits}} \\
			\hline 800 & $8 \times 10^{-9}$ & {\cite{muller2008atom}} \\
			\hline 80 & $1.7 \times 10^{-7}$ & {\cite{bodart2010cold}} \\
			\hline 2.2 & $2.3 \times 10^{-5}$ & {\cite{butts2011light}} \\
			\hline 600 & $4.2 \times 10^{-9}$ & {\cite{hu_demonstration_2013}} \\
			\hline 60 & $5 \times 10^{-8}$ & {\cite{menoret2018gravity}} \\
			\hline 2300 & $3 \times 10^{-11}$ & {\cite{dickerson2013multiaxis}} \\
			\hline 1800 & $3 \times 10^{-10}$ & {\cite{overstreet2018effective}} \\
			\hline
		\end{tabular}
	\end{table}
	
	\subsubsection{Atomic gravity gradiometer}
	Gravity gradient results from the spatial change of gravity acceleration. An atomic gravity gradiometer is an atom interferometer to measure this change in the gravitational force over space. Specifically, two atom clouds are separated by a distance $D$ in the same setup. They both measure the gravitational acceleration value at the local points simultaneously and the difference between the gravitational acceleration of the two points $\Delta g$, so the gravitational acceleration gradient $g^{\prime} = \Delta g/D$ could be obtained.
	
	Since 1998, the Kasevich group demonstrated the first atomic gravitational gradiometer based on two atomic gravimeters vertically separated by about 1 m~\cite{snadden1998measurement, mcguirk2002sensitive}, as shown in Fig.~\ref{fig:Ramaninter}(c). In this setup, the standard laser cooling technologies, including vapor cell magneto-optical trap, polarization gradient cooling, and optical pumping were used. Each cloud contained approximately $5\times 10^7$ Cs atoms at 3$\mu$K in the $F = 3$, $m_F = 0$ state as the initial state of the atomic fountain. Following the state-preparation stage, atoms were subjected to the $\pi/2$-$\pi$-$\pi/2$ Raman pulse sequence to split, reflect and recombine the matter waves. Two co-propagating Raman beams were used to minimize Doppler shifts of the Raman transition frequency, while the atomic clouds were excited by the same detection beams at the same time and the fluorescence was collected simultaneously. The accuracy of their gravity acceleration gradient measurement reached $4\times 10^{-9}g/{\rm m}$. In this experiment, they also developed a normalization algorithm called ellipse-specific fitting of sinusoidally coupled data from two gravimeters in a gradiometer configuration, which was insensitive to the main technical noise~\cite{foster2002method}.
	\begin{table}
		\centering
		\caption{Reported performance of atomic gravity gradiometers.}
		\label{tab:gravgrad}
		\begin{tabular}{|c|c|c|}
			\hline Interrogation time (ms) & Sensitivity $(g/m/ \sqrt{\mathrm{Hz}})$ & Reference \\
			\hline 60-315 & $\sim$ $3.3 \times 10^{-9}$ & {\cite{snadden1998measurement, mcguirk2002sensitive}} \\
			\hline 170 & $6 \times 10^{-9}$ & {\cite{wu2009gravity}} \\
			\hline 300 & $1 \times 10^{-8}$ & {\cite{sorrentino2014sensitivity}} \\
			\hline 330 & $6.7 \times 10^{-8}$ & {\cite{duan2014operating}} \\
			\hline
		\end{tabular}
	\end{table}
	
	In 2007, the Kasevich group obtained the Newton gravitational constant by using the atomic gravity gradiometer, as shown in Fig.~\ref{fig:Ramaninter}(d). They measured the gravitational gradient from a well-characterized Lead source mass precisely positioned between two vertically separated atomic gravimeters, so as to calculate the Newton gravitational constant. They reached the relative accuracy of about $3.2\times 10^{-3}$~\cite{fixler2007atom}. In 2008, the Tino group also used a similar setup to measure Newton's gravitational constant, but they used the denser Tungsten as the gravitational source, and obtained a relative accuracy of $4.6\times 10^{-4}$~\cite{lamporesi2008determination}, and they further improved it to $1.5\times 10^{-4}$ in 2014~\cite{rosi2014precision, sorrentino2014sensitivity}. In 2017, they improved the relative accuracy of Newton gravitational constant again~\cite{d2017canceling} and a feasible future to 10 ppm was predicted~\cite{rosi2017proposed}. Meanwhile, the Hu group at Huazhong University of Science and Technology contructed an atomic gravity gradiometer and obtained a sensitivity of {$6.7\times 10^{-8} g/m/ \sqrt{\mathrm{Hz}}$} in 2014~\cite{duan2014operating}.

	Table~\ref{tab:gravgrad} lists measured performance of some atomic gravity gradiometers reported by different groups.
	
	\subsection{Recent progresses of atom interferometer}
	It has been almost 30 years since the first demonstration of the Raman atom interferometer. Many variants and technologies have been developed to improve the sensitivity of the devices and they are used in applications of a wide range of circumstances~\cite{cronin2009optics,de2008precision,barrett2013mobile, min2015micro, bongs2019taking}. All these applications require different performance of the atom interferometers, including sensitivity, accuracy, dynamic range, stability, compactness, transportability and cost. Generally, applications in fundamental physics and metrology need atom interferometers with the highest sensitivity, while other applications not only need certain sensitivity but also some other user-related qualities. So the recent developments of Raman atom interferometers mainly focus on improving their performance in different aspects for the aimed applications.
		
	\subsubsection{Large size interferometer}
	At present, some groups are constantly pushing further to construct atomic interferometers with increasingly larger size. For instance, a 10m Raman atom interferometer has been developed by the Zhan group at Wuhan and successfully achieved a sensitivity of $2\times 10^{-7} g/\sqrt{\rm Hz}$~\cite{zhou_development_2011}. Later, the Kasevich group used a 10m atom interferometer to simultaneously measure the gravity acceleration of ${^{87}}$Rb and ${^{85}}$Rb atoms to verify the principle of equivalence in general relativity, and the sensitivity has reached $3\times 10^{-11} g/ \sqrt{\rm Hz}$~\cite{dickerson_multiaxis_2013}. Even larger projects have also been launched or proposed, such as the Matter wave-laser based Interferometer Gravitation Antenna (MIGA) project started in 2013~\cite{canuel2018exploring}, the Atomic Interferometric Observatory and Network (AION) project designed for unprecedentedly high sensitivity to detect gravitational wave and to search for dark matter~\cite{badurina2020aion}, and the Zhaoshan long-baseline Atom Interferometer Gravitation Antenna (ZAIGA) project proposed~\cite{zhan2019zaiga}. Being proposed in 2019, the ZAIGA project plans to build a underground laser-linked interferometer facility in a 300-meter vertical tunnel for atom fountain and atom clocks, aiming to explore fundamental physics of gravitation and related problems~\cite{zhan2019zaiga}.
	
	\subsubsection{High momentum transfer}
	To increase the enclosed area of an atom interferometer for high sensitivity, an alternative way is using laser pulses to transfer sufficiently large lateral recoil momentum to the atoms. In 1991, the standard $\pi/2$-$\pi$-$\pi/2$ Raman pulses only transferred $2\hbar k$ momentum to the atoms~\cite{kasevich1991atomic}. In 2008, the Chu group tried Bragg scattering to transfer $24\hbar k$ momentum to the atoms, increasing the phase shift 12-fold for the Mach-Zehnder atom interferometer~\cite{muller_atom_2008}. In 2011, the Kasevich group used a sequential multi-photon Bragg diffractions to transfer up to $102\hbar k$ momentum to the atoms~\cite{chiow_$102ensuremathhbark$_2011}. There seems no impediments to scaling this method to even larger momentum transfer, perhaps in excess of $1000 \hbar k$. This technique could be very promising to achieve even higher sensitivity of the current atom interferometers and to construct compact devices with excellent performance.
	
	\subsubsection{Portable atom interferometer}
	In addition to fundamental physics, atom interferometers have demonstrated many other potential applications such as inertial navigation, gravity detection and mineral surveys due to their high sensitivity. In these scenarios, a compact and portable system for field applications is very much desired. In the past decade, massive effort has been devoted to different portable systems in many groups worldwide~\cite{barrett2013mobile}. For instance, the size of the atomic gyroscope based on dual interferometers from the Kasevich group was smaller than $0.6 \times 0.6 \times 1 \ {\rm m}^3$, while reached a sensitivity of $7\times 10^{-6}({\rm rad/s})/ \sqrt{\rm Hz}$~\cite{takase2008precision}. Their compact gravity gradiometer was smaller than \SI{1}{\metre\cubed}, a sensitivity was $6\times 10^{-9}{g/{\rm m}}/\sqrt{\rm Hz}$~\cite{wu2009gravity}. P. Bouyer{\it et al.} have also been trying to construct smaller atom interferometers~\cite{le2008limits}. The sensor head of their gravimeter with the unique single-beam pyramidal magneto-optical trap was only \SI{0.4}{\metre}, and a $ 1.4\times 10^{-8}g/ \sqrt{\rm Hz}$ sensitivity had be obtained. Their dual interferometer gyroscope with total length of \SI{0.9}{\metre} have reached a sensitivity of $2\times 10^{-4}({\rm rad/s})/ \sqrt{\rm Hz}$~\cite{muller_compact_2009}. 
	Portable atom interferometers have being drawing special attention in the last three years. In 2017, the Zhan group at Wuhan Institute of Physics and Mathematics of the Chinese Academy of Sciences demonstrated a $^{85}$Rb atom gravimeter (WAG-H5-1) and obtained a sensitivity of {$30 \mu {\rm Gal}/\sqrt{\rm Hz}$} and a stability of $1 \mu {\rm Gal}@4000 {\rm}s$ ~\cite{huang2019accuracy}. In 2018, the Li group at National Institute of Metrology constructed a Raman atom gravimeter (NIM-AGRb-1) and obtained a sensitivity of 44~$\mu {\rm Gal}/\sqrt{\rm Hz}$  ~\cite{wang2018shift}. In 2019, the Hu group at Huazhong University of Science and Technology reported a portable gravimeter with a sensitivity of 53~$\mu {\rm Gal}/\sqrt{\rm Hz}$~\cite{luo2019compact}. In the same year, the Lin group at Zhejiang University of Technology realized a sensitivity of 300~$\mu {\rm Gal}/\sqrt{\rm Hz}$ in their Raman atom gravimeter~\cite{fu2019new}. Recently, the Chen group at University of Science and Technology of China completed a Raman atom gravimeter (USTC-AG02) and achieved a sensitivity of 35~$\mu{\rm  Gal}/\sqrt{\rm Hz}$ ~\cite{xie2020tilt}.
		\begin{figure}
			\centering
			\includegraphics[width=9cm]{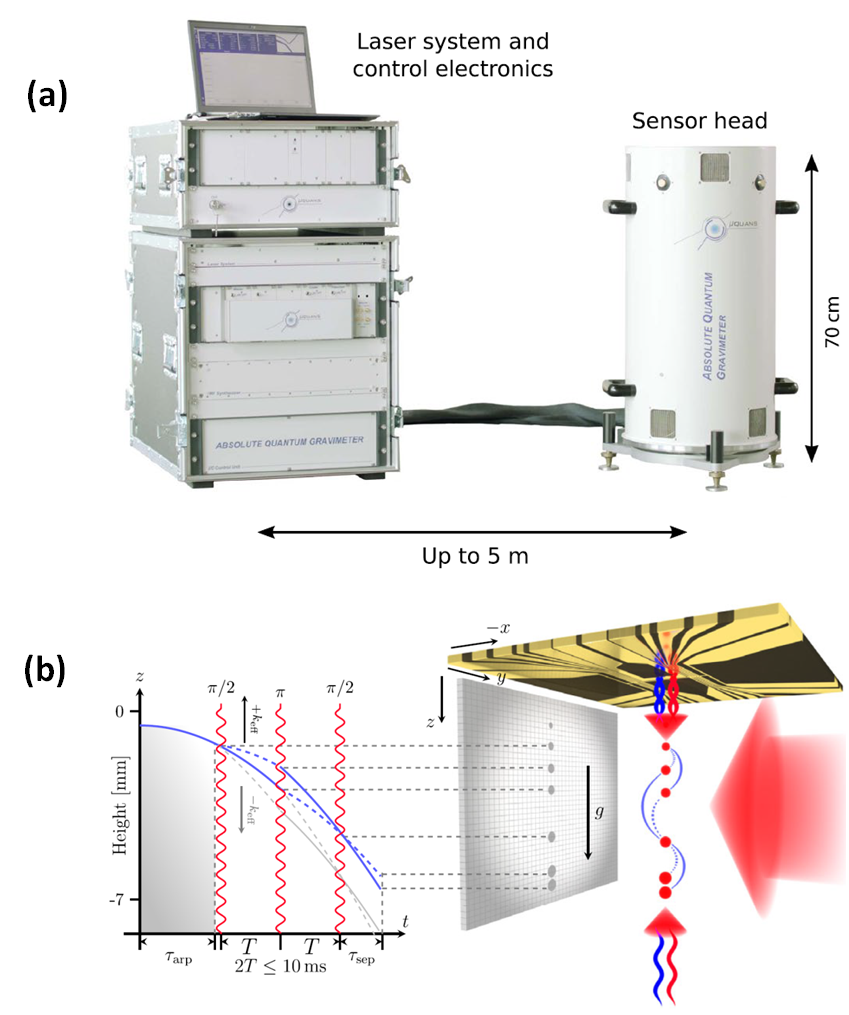}
			\caption{New variants of Raman atom interferometers. (a) A transportable absolute quantum gravimeter with a long-term stability below $1\times 10^{-9}g$~\cite{menoret2018gravity}. (b) An atom-chip fountain gravimeter based on a freely falling Bose-Einstein condensates from an atomic chip~\cite{abend2016atom}.}
			\label{fig:commercial}
		\end{figure}
		%
		
	\subsubsection{Commercial atom interferometer}
	At present, more atom interferometers are moving out of the laboratories to be used outdoors. Compact and portable atomic gravimeters are commercially available from companies like MUQUANS and AOSENSE, which are founded by the PIs of the leading research groups. The sizes of these commercial gravimeters are typically shorter than \SI{1}{\metre} and smaller than \SI{0.3}{\metre} in diameter. As shown in Fig.~\ref{fig:commercial}(a), the device from MUQUANS reached a long-term stability of better than $10 {\rm nm}/{\rm s}^2$, which is comparable to commercial falling corner-cube instruments~\cite{menoret2018gravity}.
		
	\subsubsection{Multi-axis interferometer}
	When an atom interferometer is used for inertial navigation, the accelerations and angular velocities in all three directions are needed. However, it is currently impractical to construct a system to contain six atom interferometers in a reasonably small space at the same time. {The Landragin group} developed an unique six-axis atom interferometer to provide a full inertial base~\cite{canuel_six-axis_2006, gauguet_characterization_2009}. This design used two counter-propagating cold-atom clouds that were launched in curved parabolic trajectories and three single Raman beam pairs pulsed in orthogonal directions, so that rotations about the three axes and accelerations along the three directions were accessible simultaneously. This one-vacuum-system atom interferometer realized six-axis measurement and reached a sensitivity of $1.4\times 10^{-7}({\rm rad/s})$ to rotation and $6.4\times 10^{-7}{\rm m/s}^2$ to acceleration after 600 seconds of averaging time. In addition, the atom interferometer with point source atoms in the Kasevich group~\cite{dickerson_multiaxis_2013} used the velocity dependent Coriolis forces and spatially resolved detection to measure the gravity acceleration and the angular velocity simultaneously. A single shot sensitivity of $6.7\times 10^{-12}g$ and $2.0\times 10^{-7}({\rm rad/s})$ was demonstrated. In 2019, a very compact design of single-source multi-axis atom interferometer in a centimeter-scale cell showed a sensitivity of $1.6\times 10^{-5}/\sqrt{\rm Hz}$ and $5.7\times 10^{-5}({\rm rad/s})$~\cite{chen2019single}.
		
	\subsubsection{Chip-scale interferometer}
	Atom Chips are micro-fabricated wires patterned on the surface to confine, control and manipulate cold atoms~\cite{keil2016fifteen}, which have been used to control and guide atoms coherently~\cite{wang_atom_2005,bucker2011twin,jo2007long}. A freely falling Bose-Einstein condensate from an atom chip was used as an atomic gravimeter and an accuracy of $1.7\times 10^{-7}g$ was obtained~\cite{abend2016atom}, as shown in Fig.~\ref{fig:commercial}(b). These technologies will become more important for developing compact matter-wave interferometers for certain applications.
		
	\subsubsection{Atom source}
	The atom number or beam flux changes the signal-to-noise ratio of the measured inference fringes and determines the highest sensitivity attainable of an atom interferometer. So it is always favorable to improve the design of atom sources and laser cooling techniques, therefore enhance the detecting signal and obtain higher sensitivity~\cite{muller_versatile_2007}. In a cold atom interferometer, atoms are maintained at low temperature such that the spatially thermal expansion and wavefront aberrations of the matter wave are both reduced, which helps to improve the sensitivity~\cite{kerman_beyond_2000, dickerson_multiaxis_2013, karcher2018improving}.
		
	\subsubsection{Ion-based interferometer}
	Different from the atom-based quantum sensor, a trapped ion quantum sensor possesses some practical advantages owing to due to its compact setup, simple preparation and large scale factor~\cite{campbell2017rotation}. It is an ideal testing platform for future compact and portable devices in many applications.
		
	In 2017, trapped-ion was proposed as a rotation sensor via matter-wave Sagnac interferometry~\cite{campbell2017rotation}. This protocol measures rotation with a single ion that hosts a qubit with its internal states $\ket{\uparrow}$ or $\ket{\downarrow }$. This  interferometer encloses effectively a large area in a compact apparatus through repeated round-trips in a Sagnac geometry. Unlike the Raman pulses used in atom interferometer, $\pi/2$ pulses and spin-dependent kicks can close the paths of the ion-based gyroscope to access the rotation induced phase shift.
		
	The ion-based gyroscope is currently witnessing its early development in laboratories. Imperfections in trap potential have been identified to be responsible for various unwanted systematic effects, such as common misalignment of imparted momenta and relative misalignment of imparted momenta, which can affect both the interferometer phase and visibility. In 2019, A. West~\cite{west2019systematic} examined these systematic effects in a trapped ion-based matter-wave interferometer for rotation sensing in particular, and found that good control of the trapping potential can make interferometer insensitive to experimental imperfections. In the same year, E. Urban {\it et al.}~\cite{urban2019coherent} demonstrated coherent control of quantum rotor of two-ion Coulomb crystal in a circularly symmetric potential. It would inspire more experiments of trapped ions for quantum interferometry and sensing.
		
	\section{Summary}
	For the precision measurements with cold atoms and ions, higher precision and smaller device are the two most important directions for future developments. The pursuit for higher precision is a forever topic in metrology and fundamental physics, which requires the improving of state-of-the-art technologies or inventing new variants of clocks and interferometers. For example, to improve the fundamental performance of optical atomic clocks, one important approach is to look for new clock transitions with higher frequencies. One possible choice is a nuclear transition of $^{229}$Th, which has been recently identified~\cite{masuda2019x, seiferle2019energy}. Such a transition is suggested to have the potential to improve the clock stability by about five orders of magnitude. For cold atom magnetometers, novel designs are always welcome to achieve better performance, while at the same time to test novel quantum technologies for quantum sensing in general, such as spin squeezing and entanglement~\cite{muessel2014scalable, ockeloen2013quantum}. Regarding to the atom interferometers,  constructing larger size interferometers~\cite{canuel2018exploring, zhan2019zaiga} or using them in special environments, e.g. in micro-gravity in space~\cite{tino2019sage}, will further push the sensitivity limit for fundamental physics and metrology.  
		
	An equally important and simultaneously challenging task is to build smaller and transportable devices by simplifying and minifying equipments. Transportable optical atomic clocks will still be a valuable directions~\cite{bize2005cold} as this kind of clock has been demonstrated for geodesy application~\cite{grotti2018geodesy}. Additionally, miniature setups also find their special applications in certain fields, and efforts on developing of smaller magnetometers and interferometers would never be overemphasized~\cite{eliasson2019spatially, barrett2013mobile}.
		
	In summary, we briefly introduce in this short review some recent progresses, perspectives and challenges in the field of precision measurement, paying special attention to matter-wave interferometers based on cold atoms and trapped ions. We summarize some recent research advances in this field towards the development of better measuring equipments for frequency, magnetic field, and inertial forces. These new progresses not only pave a route to achieve a higher sensitivity and a stronger stability, but also facilitate the specialized applications in various circumstances ranging from fundamental sciences to commercial industries, where a precise measurement of quantities is in need. Finally, we would like to conclude this review with a famous quote by Johannes Kepler:
	``{\it Just as the eye was made to see colors, and the ear to hear sounds, so the human mind was made to understand, not whatever you please, but quantity.}"
		
	\addcontentsline{toc}{chapter}{Acknowledgment}
	\section*{Acknowledgment}
	We would like to thank Meng Shi and Jian Cao for discussions on theoretical and experimental issues. This work is supported by the National Natural Science Foundation of China (Grant No.~11522436,~11774425,~11704408,~91836106), the Beijing Natural Science Foundation (Grant No.~Z180013) and the Joint fund of the Ministry of Education (Grant No.~6141A020333xx), and the Research Funds of Renmin University of China (Grant No.~16XNLQ03,~18XNLQ15).
		

		
	\end{document}